\documentclass[prb,
twocolumn,
superscriptaddress,showpacs,amsmath,amssymb]{revtex4}

\begin{document}

\title{Combined Lorentz symmetry: lessons from superfluid $^3$He}

\author{G.E.~Volovik}
\affiliation{Low Temperature Laboratory, Aalto University,  P.O. Box 15100, FI-00076 Aalto, Finland}
\affiliation{Landau Institute for Theoretical Physics, acad. Semyonov av., 1a, 142432,
Chernogolovka, Russia}

\date{\today}

\begin{abstract}
{We consider the possibility of the scenario in which the  $P$, $T$ and Lorentz symmetry of the relativistic quantum vacuum are all the combined symmetries. These symmetries emerge as a result of the symmetry breaking of the more fundamental  $P$, $T$ and Lorentz symmetries of the original vacuum, which is invariant under separate groups of the coordinate transformations and spin rotations. The condensed matter vacua (ground states) suggest two possible scenarios of the origin of the combined Lorentz symmetry, both are realized in the superfluid phases of liquid $^3$He:
the $^3$He-A scenario and the  $^3$He-B scenario. In these scenarios  the gravitational tetrads are considered as the order parameter of the symmetry breaking in the quantum vacuum. The $^3$He-B scenarios applied to the Minkowski vacuum leads to the continuous degeneracy of the Minkowski vacuum with respect to the $O(3,1)$ spin rotations. The symmetry breaking leads to the corresponding topological objects, which appear due to the nontrivial topology of the manifold of the degenerate Minkowski vacua, such as torsion strings. The  4-fold degeneracy of the Minkowski vacuum with respect to  discrete $P$ and $T$ symmetries suggests that the Weyl fermions are described by four different tetrad fields: the tetrad for the left-handed fermions, the tetrad for the right-handed fermions, and the tetrads for their antiparticles. This may lead to the gravity with several metric fields, so that  the parity violation may lead to the breaking of equivalence principle. Finally we considered the application of the gravitational tetrads for the solution of the cosmological constant problem.
}
\end{abstract}
\pacs{
}

\maketitle

\section{Introduction}

Topological superfluid phases of liquid $^3$He provide many connections with the Standard Model of particle physics and gravity.\cite{Volovik2021,Volovik2003} Here we consider the symmetries and degeneracy of the Minkowski vacuum from the point of vies of the symmetry breaking in the B-phase of superfluid $^3$He.\cite{VollhardtWolfle} 

In the Standard Model, the discrete  symmetries $P$ and $T$ are broken. It is possible that these symmetries are restored at some ultraviolet  energy scale, which is below the Planck scale $E_{\rm UV} \ll E_{\rm P}$. However, even if $T$ and $P$ are restored above $E_{\rm UV}$, they still have the signatures of the broken symmetry, since each of them represents the combined symmetry. The fermionic action is not invariant under the pure transformations of coordinates, $P_{\rm c} {\bf r} = -  {\bf r}$ and $T_{\rm c} t = - t$. It becomes invariant only if these coordinate transformations are accompanied by the corresponding transformations of Dirac or Weyl spinors, $P_{\rm s}$ and $T_{\rm s}$. This means that each of these two discrete symmetry operations is the product of two operations,
 $P=P_{\rm c}  P_{\rm s}$ and  $T=T_{\rm c} T_{\rm s}$. Since $P_{\rm c}$ and $T_{\rm c}$ are not the symmetries of the  fermionic quantum vacuum even in the high-energy scale, it is natural to suggest, that the Minkowski vacuum is degenerate with respect to these discrete symmetries.  Such degeneracy of the quantum vacuum has been suggested by Vergeles.\cite{Vergeles2021} 
 
 The natural extension of this idea is the application to  the continuous Lorentz symmetry $L$. The fermionic action lacks the separate symmetries, such as symmetry with respect to the only coordinate transformations $L_{\rm c}$, but is invariant under the combined operation, $L=L_{\rm c} L_{\rm s}$, i.e. when the space rotations are accompanied to the spin rotations of the Dirac or Weyl spinors. One may suggest the symmetry breaking pattern  $L_{\rm c}  \times L_{\rm s}  \rightarrow L$, which  follows from the consideration of similar 
symmetry breaking in superfluid $^3$He-B.\cite{Volovik1990}  The original vacuum of normal liquid $^3$He is invariant and separate spin and orbital (coordinate) rotations,  $G=SO(3)_{\rm s}  \times SO(3)_{\rm c}$. In the B-phase this symmetry is reduced to the diagonal group 
$G \rightarrow H = SO(3)_J$ of the combined rotations.  Such symmetry breaking is known as the broken relative symmetry,\cite{Leggett1973} the vacuum becomes degenerate with respect to the separate coordinate or spin rotations. Extension to the Lorentz groups suggests the continuous degeneracy of the Minkowski quantum vacuum. 

The same take place with the discrete parity symmetry, $P_{\rm c} \times P_{\rm s} \rightarrow P_{\rm c}P_{\rm s}$, which leads to the additional discrete degeneracy of the $^3$He-B and correspondingly of the Minkowski vacuum. Here we consider some consequences of this symmetry breaking pattern in  the Minkowski vacuum.
The paper is organized as follows. 

In Sec. \ref{Aphase} the topological scenario of  emergent combined symmetry is discussed, which follows from the analogy with the chiral superfluid $^3$He-A with emergent Weyl fermions.

In Sec. \ref{Bphase} the B-phase symmetry breaking is discussed. The spontaneous breaking of the separate rotational symmetries  is demonstrated together with the spontaneous breaking of discrete symmetry $P_{\rm c} \times P_{\rm s} \rightarrow P=P_{\rm c}  P_{\rm s}$.

In Sec. \ref{Relativistic} the  symmetry breaking scenario, which takes place in $^3$He-B, is extended to the relativistic Minkowski vacuum. 
Two independent Lorentz groups of the coordinate and spin rotations are broken to the diagonal subgroup of the degenerate Minkowski vacuum. The broken discrete symmetries and the corresponding degeneracy of the  Minkowski vacuum are also discussed.

The topological defects arising from the breaking of relative symmetry, both in  $^3$He-B and in the relativistic vacuum, are discussed in Sec. \ref{defects}. They are obtained from the homotopy groups of the manifold of the degenerate states, $\pi_n(G/H)$.  In Sec. \ref{TorsionString} the particular topological object emerging due to the degeneracy of Minkowski vacuum is discussed -- the torsion string.

 In Sec. \ref{MagneticMomentSec} the mirror reflected vacuum is discussed, which is obtained by the discrete symmetry operation acting on our Minkowski vacuum. We discuss the magnetic moment of electron in the mirror reflected vacuum. 

In Sec. \ref{MultipleTetrads} we discuss the fermionic action for Weyl fermions instead of the Dirac fermions, which are the secondary composite objects. The broken discrete symmetries and the corresponding 4-fold degeneracy of the Minkowski vacuum suggest that the Weyl fermions can be described by four tetrad fields: each for left-handed and right-handed fermions and for their antiparticles. 
Though the considerations in this paper are mainly restricted by the Minkowski vacua and their degeneracy, 

in Sec. \ref{bimetric}  we consider the possibe bi-metric gravity, where different metric can be composed from different tetrad fields.

In Secs. \ref{CC} and \ref{Thermodynamics} we considered the possible application of the tetrad fields for the solution of the cosmological constant problem.

\section{Topological A-phase scenario:  emergent combined symmetry}
\label{Aphase}

The condensed matter vacua suggest two possible scenarios of the origin of the combined Lorentz symmetry, which are realized in the superfluid phases of liquid $^3$He:
the $^3$He-A scenario and the  $^3$He-B scenario. The latter scenario corresponds to the phase transition with spontaneous breaking of separate Lorentz symmetries $L_{\rm c}$ and $L_{\rm s}$, see Sec. \ref{Bphase}.

$^3$He-A is the chiral superfluid, which has the topologically protected Weyl points in the quasiparticle spectrum.\cite{Volovik2003}
In the vicinity of the Weyl point, quasiparticles behave as Weyl fermions moving in the effective dynamical gauge and tetrad fields.
The combined Lorentz symmetry emerges together with the tetrads and Weyl fermions. 
In this scenario, the separate Lorentz symmetries, which independently act on coordinates and on fermions, are absent, since on the microscopic atomic scale (the analog of Planckian scale in quantum liquids)  the physics is Galilean.

The effective gravity for Weyl fermions, provides an example of degeneracy of quantum vacuum with respect to the discrete symmetry -- the $P$ symmetry. The quantum transition is possible in the A-phase, at which some components of the tetrad change sign.\cite{NissinenVolovik2018} At the transition the tetrad is degenerate, and two Weyl points of the A-phase transform to the Dirac nodal line in the quasiparticle spectrum of the polar phase.\cite{Autti2020} The superfluid polar phase, which is time reversal invariant,
connects two degenerate vacua with opposite chiralities -- analogs of spacetime and antispacetime.

\section{B-phase scenario:  combined symmetry from broken symmetry}
\label{Bphase}

The $^3$He-B is the fully gapped superfluid, which is obtained in the following scenario of the spontaneous symmetry breaking.\cite{VollhardtWolfle} 
The original (non-superfluid) state of liquid $^3$He can be described by the following relevant Hamiltonian:
\begin{eqnarray}
{\cal H}-\mu {\cal N}=
\sum_{{\bf p }\alpha}\left({p^2\over
2m}-\mu\right)a^\dagger_{{\bf p }\alpha}a_{{\bf p }\alpha}-
\nonumber 
\\
-\frac{\lambda}{2}
 \sum_{{\bf p'}\alpha{\bf p} \beta}({\bf p}'\cdot{\bf p}) \,
 a^\dagger_{-{\bf p}'\beta}a^\dagger_{{\bf p}'\alpha}  
 a_{{\bf p}\alpha}a_{-{\bf p}\beta} \,.
\label{NormalLiquid}
\end{eqnarray}
Here $\alpha= (\uparrow, \downarrow)$ and  $\beta= (\uparrow, \downarrow)$ denote the spin projections of an atom of $^3$He, and  $\lambda$ is the small parameter of the relevant  interaction. Here we ignored the $U(1)$ symmetry breaking, which leads to the phase factor and makes the order parameter complex.\cite{Volovik1990}

 The symmetric  state of the normal (non-superfluid)  liquid $^3$He has the symmetry of the Hamiltonian (\ref{NormalLiquid}). It is symmetric under  $SO(3)_{\rm c}$ group of the orbital rotations and, if the tiny spin-orbit interaction is neglected, it is also symmetric under rotations $SO(3)_{\rm s}$ in spin space.  
 
 In the superfluid state, the bilinear form $a_{{\bf p}\alpha}a_{-{\bf p}\beta}$
acquires a non-zero vacuum expectation value. As a result, for $^3$He-B the Hamiltonian (\ref{NormalLiquid}) transforms to the Hamiltonian with 4-component Bogolyubov-Nambu spinor $\chi=(\Psi,\Psi^\dagger)$ with the spectrum:
\begin{equation}
H_{\bf p} =\left({p^2\over 2m} - \mu\right)\tau_3 + e^i_a \tau_1\sigma^a p_i \,.
\label{3HeB}
\end{equation}
Here $\tau_1$, $\tau_3$ and $\sigma^a$ with $a=1,2,3$ are the $2\times 2$ Pauli matrices, and the matrix $e^i_a $ is the order parameter, which is obtained as vacuum expectation value of the bilinear form: 
\begin{equation}
\lambda \left<\sum_{\bf p}{\bf p} a_{{\bf p}\alpha}a_{-{\bf p}\beta}\right>_{\rm vac}=  
{\bf e}_{a}  (i\sigma ^{a}\sigma ^{(2)})_{\alpha\beta} \,.
\label{AnomalousAverageGeneralSpin}
\end{equation}
The Hamiltonian (\ref{3HeB}) is the analog of the Dirac Hamiltonian, and in the limit $m\rightarrow \infty$ it approaches the Hamiltonian for the Dirac spinors\cite{SilaevVolovik2010} with mass $M=\mu$:
\begin{equation}
H_{\bf p} =  - \mu\tau_3 + e^i_a \tau_1\sigma^a p_i \,,
\label{3HeBrel}
\end{equation}
and the matrices $\tau_3$ and $ \tau_1\sigma^a$ become the corresponding $\gamma$ matrices.

The order parameter $e_a^i$  breaks both rotational symmetries of the original Hamiltonian, since it transforms under $SO(3)_{\rm c}$ group of the coordinate rotations and under  $SO(3)_{\rm s}$ group of the spin rotations. This means that the vacuum in $^3$He-B is degenerate with respect to spin or to orbital rotations. 
 But the symmetry $SO(3)_J$ under the combined spin and orbital rotations is obeyed, i.e. the spin rotation can be compensated by the opposite orbital rotation. That is why the symmetry breaking pattern is:
\begin{equation}
SO(3)_{\rm c}\times SO(3)_{\rm s}\rightarrow SO(3)_J \,.
\label{BrokenSymmetry}
\end{equation}
Such symmetry breaking is called the broken relative symmetry.\cite{Leggett1973}

In $^3$He-B, the order parameter matrix $e_a^i$ plays the role of the triad field in the emerging $3+0$ effective gravity. That is why the $SO(3)_{\rm c}$ group of orbital rotations  plays the role of the group  of coordinate transformations, while $SO(3)_{\rm s}$ group of spin rotations  plays the role of the group of the transformations of triads in spin space.

Let us also consider the discrete symmetries on example of parity $P$, which is broken in $^3$He-B. The original non-relativistic Hamiltonian (\ref{NormalLiquid}) for the normal liquid is invariant both under the coordinate transformation $P_{\rm c}\Psi({\bf r},t) =\Psi(-{\bf r},t)$, and under transformation $P_{\rm s}$ in the spin space, since the non-relativistic fermions are invariant under parity transformation, $P_{\rm s} \Psi P_{\rm s} =\Psi$.

In the broken symmetry state each of these two discrete symmetries is broken, and the Hamiltonian (\ref{3HeB}) for $^3$He-B is invariant only under combined symmetry:   $P=P_{\rm c}  P_{\rm s}$, where  $P_{\rm c} \chi ({\bf r},t) =\chi(-{\bf r},t)$  is the coordinate inversion and  $P_{\rm s} \chi =\tau_3\chi\tau_3$ is the inversion in spin space. 
Taking parity into account the symmetry breaking scheme in $^3$He-B becomes
\begin{equation}
G=SO(3)_{\rm c}\times P_{\rm c} \times SO(3)_{\rm s} \times  P_{\rm s} \rightarrow H=SO(3)_J \times P \,.
\label{SymmetryBreakingB}
\end{equation}
Note again, that we ignore the $U(1)$ symmetry breaking  in superfluid $^3$He-B, and consider only the real components of the order parameter. We also did not consider here the time reversal symmetry, since it is not broken in $^3$He-B.

\section{Extension to relativistic physics}
\label{Relativistic}

It is natural to extend the symmetry breaking scheme presented in Eq.(\ref{BrokenSymmetry})  to the 3+1 case of Minkowski vacuum.\cite{Volovik1990}  The analogy from $^3$He-B suggests that there is the original deep quantum vacuum, which is symmetric with respect to the group $G$, which contains two independent transformations: the vacuum has the symmetry under the  Lorentz space-time rotations (which can be extended to the group of coordinate transformations), and the symmetry under the Lorentz  3+1 transformations of the fermionic spins. This vacuum experiences the symmetry breaking of these two Lorentz symmetries to their diagonal subgroup: 
\begin{equation}
 L_{\rm c} \times L_{\rm s} \rightarrow L_{J} \,.
\label{BrokenSymmetry4}
\end{equation}
The order parameter of this symmetry breaking is the matrix $e^\mu_a$, which plays the role of the tetrad  of the (maybe emergent) $3+1$ gravity. 
The Minkowski vacuum becomes degenerate with respect to each of the two rotations, but remains invariant under the combined Lorentz transformation. The latter represents the rest symmetry $H$ of the present degenerate vacuum.

Simultaneously, the original fermions acquire the form of the Dirac or Weyl fermions, which obey the Weyl or Dirac action containing the tetrad field.  
The gravity interacting with fermions is described by the Einstein-Cartan-Sciama-Kibble theory and its possible extensions.
The typical action for Dirac fermions is:
\begin{equation}
U_\Psi= \frac{1}{6}e_{abcd}\int \Theta^a\wedge e^b \wedge e^c\wedge e^d
\,,
\label{Action1}
\end{equation} 
\begin{equation}
 \Theta^a = \frac{i}{2} \left[ \bar\Psi \gamma^a D_\mu \Psi - D_\mu \bar\Psi\gamma^a \Psi\right] dx^\mu
\,,
\label{Action2}
\end{equation}
where the spin indices are raised with $\eta^{ab}$, the flat Minkowski metric.
The state of the broken symmetry vacuum and the corresponding Dirac or Weyl action remain invariant only if the transformation of the coordinate $x^\mu$ is accompanied by the proper rotations of the tetrads $e^a_\mu$  in the spin space.
These combined transformation may also include the operations of discrete symmetries. For example, the combined $PT$ symmetry is the discrete 
 coordinate transformation $(PT)_{\rm c}  x^\mu =-x^\mu$, which is accompanied by the corresponding  transformation of the tetrads 
 $(PT)_{\rm s}  e_\mu^a =- e_\mu^a$.  The action is invariant under transformation $PT= (PT)_{\rm c}   (PT)_{\rm s}$, but is not invariant under the separate symmetry operation  $(PT)_{\rm s}$, which  transforms $e^a$ and $-e^a$. This $(PT)_{\rm s}$ symmetry is broken, and thus the vacua with $e^a$ and $-e^a$ are different.\cite{Vergeles2021}
  Just such scenario of the symmetry breaking takes place in the theories, where the tetrad field emerges as bilinear form of the fermionic fields.\cite{Diakonov2011,Diakonov2012,ObukhovHehl2012,Wetterich2003,Wetterich2012,Wetterich2013}  

Taking into account both the discrete and continuous symmetries, the natural extension of the symmetry breaking in Eq.(\ref{SymmetryBreakingB})
to the relativistic physics will be 
\begin{eqnarray}
G=L_{\rm c}\times P_{\rm c}  \times T_{\rm c}\times L_{\rm s} \times  P_{\rm s}  \times T_{\rm s} \rightarrow H \,,
\label{SymmetryBreakingUniverse1}
\\
 H= L\times P \times T
\,,
\label{SymmetryBreakingUniverse2}
\\
L=L_{\rm c}L_{\rm s} \,\,,\,\, P= P_{\rm c}P_{\rm s}  \,\,,\,\, T= T_{\rm c}T_{\rm s} \,.
\label{SymmetryBreakingUniverse3}
\end{eqnarray}

The simplest model illustrating this symmetry breaking pattern $G\rightarrow H$ has the following action on the  microscopic scale:\cite{Wetterich2003}
 \begin{equation}
U_{\rm micro}= \frac{1}{24}  e_{abcd}\int  \Theta^a\wedge \Theta^b \wedge \Theta^c\wedge \Theta^d \,.
\label{Initial}
\end{equation}
This action has the full symmetry $G$.
The symmetry breaking is represented by the appearance of the Bogoliubov quasi-average:
 \begin{equation}
e^a =\left< \Theta^a\right> \,.
\label{quasiaveragel}
\end{equation}
This order parameter $e^a_\mu$ plays the role of the emerging tetrad field. The zeroth term in the expansion gives the gravitational tern corresponding to the cosmological constant:
 \begin{equation}
U_0= \frac{1}{24}  e_{abcd}\int e^a\wedge e^b \wedge e^c \wedge e^d \,,
\label{ZerothOrderl}
\end{equation}
while the first order term in expansion produces the action (\ref{Action1}) for fermions with  rest symmetry $H$.

 \section{Degenerate Minkowski vacua and topological objects}
 \label{defects}

In $^3$He-B, the symmetry breaking scheme in Eq.(\ref{SymmetryBreakingB}) leads to several topological objects -- the topologically stable configurations of the order parameter $e^i_a$. They are described by the homotopy groups of the space $R=G/H$ -- the manifold of degenerate states: 
\begin{eqnarray}
\pi_n(R)=\pi_n(G/H)=\pi_n(SO(3)\times Z_2) \,.
\label{nB2}
\end{eqnarray}
The group $\pi_0(R)=Z_2$ gives rise to the domain walls discussed in Ref.\cite{SalomaaVolovik1988}.
When the  $U(1)$ symmetry breaking is taken into account, these domain walls become the analogs of the Kibble-Lazarides-Shafi walls  bounded by the Alice strings \cite{Kibble1982} (half-quantum vortices).\cite{SalomaaVolovik1988,Makinen2019,Kuang2020}

The fundamental group of the manifold of the degenerate vacuum states in $^3$He-B is $\pi_1(G/H)=Z_2$.
The corresponding topological objects are the $Z_2$ spin-orbital vortices.\cite{SalomaaVolovik1987}
They have been identified in NMR experiments on $^3$He-B.\cite{Kondo1992}

Similar topological objects may exist in the $3+1$ tetrad gravity.
As follows from Eqs.(\ref{SymmetryBreakingUniverse1}) and (\ref{SymmetryBreakingUniverse2})
the homotopy groups of the manifold of the degenerate Minkowski vacua are
\begin{eqnarray}
 \pi_n(G/H)=\pi_n(O(3,1))
 =\pi_n(SO(3,1)\times Z_2\times Z_2) \,,
\label{nL}
\end{eqnarray}
where we took into account the broken discrete symmetries: parity $P_{\rm c}  \times  P_{\rm s}\rightarrow P$ and time reversal $T_{\rm c}  \times  T_{\rm s} \rightarrow T$.

The group $\pi_0(G/H)=Z_2\times Z_2$ gives rise to three types of cosmic domain walls, in which either the space components of tetrads  change sign, or time component changes sign, or the whole tetrad changes sign.

The fundamental group of the manifold of the Minkowski vacuum states is $\pi_1(G/H)=Z_2$. It is the same as in  $^3$He-B, where it gives rise to  the spin-orbital vortex. Its analog in Minkowski vacuum is the torsion string, in which the curvature has $\delta$-function singularity on the axis, see Sec. \ref{TorsionString}.

\section{Torsion string in the broken symmetry vacuum}
\label{TorsionString}

Here we consider the torsion string following discussion with Jacobson \cite{Jacobson}. Let us consider  the tetrad 1-forms $e^a$ and
the SO(3,1) spin connection 1-form $\omega^a_b$ as independent variables. The action for pure GR without matter is
\begin{equation}
\int e^a\wedge e^b\wedge R^{cd} \epsilon_{abcd}.
\label{action}
\end{equation}
 The equations of motion obtained by variation over $\omega^a_b$ and $e^a$  are correspondingly
\begin{equation}
D(e^{[a}\wedge e^{b]})=0 \,,
\label{equation1}
\end{equation}
\begin{equation}
 e^b\wedge R^{cd}\epsilon_{abcd}=0 \,.
\label{equation2}
\end{equation}

We are interested in such the torsion strings, where tetrads have the following axisymmetric form:
\begin{equation}
e^0=dt,\qquad e^3=dz, \qquad e^1=dr,\qquad e^2=r\,
d\phi.
\end{equation}
This structure corresponds to the vortex in chiral $^3$He-A and also to the spin vortex in $^3$He-B -- the $Z_2$ defect in the order parameter matrix $R_a^i$. 

The metric does not feel the topology of the space dependent tetrad field, the space part of the metric is:
\begin{equation} 
g^{ik}=- \delta^{ik} \,.
\label{MetricInString}
\end{equation}  
Thus the Riemann curvature, calculated directly from the metric is zero. However, the curvature 2-form  $R^a_b$  calculated from the tetrads has singularity. The first equation of motion, Eq.(\ref{equation1}), has solution for the spin connection
\begin{equation}
\omega^0_a=0,\qquad \omega^3_a=0,\qquad \omega^1{}_2 = \frac{1}{r} e^2,
\end{equation} 
 which gives the singularity in the curvature on the vortex axis:
\begin{equation}
R_{12ik} = e_{ikl}{\hat z}^l \delta_2({\bf r})\,.
\label{RiemannCurvature}
\end{equation}

The other topological objects in tetrad gravity are the topological instantons. \cite{HansonRegge,AuriaRegge}
In principle, the Big Bang can be also considered as the topological defect, in which the tetrads change sign.\cite{Sakharov1980,Turok2018,Turok2021}

\section{Sign of gravitational tetrads and magnetic moment of electron}
\label{MagneticMomentSec}

Let us consider some consequences of the broken discrete symmetries, such as $(PT)_{\rm c} \times  (PT)_{\rm s} \rightarrow PT$.
We start with the anomalous magnetic dipole moment of the electron, which is determined by the following 3+1 term in the Lagrangian:
 \begin{equation}
L \propto  i e_a^\mu e_b^\nu \bar\psi    (\gamma^a \gamma^b - \gamma^b \gamma^a) \psi F_{\mu\nu} \,,
 \label{Action}
\end{equation}
where $F_{\mu\nu}$ is the electromagnetic field.
We consider the magnetic field ${\bf B} \parallel \hat{\bf z}$. The projection on $z$ axis of the magnetic moment of electron, which is the variation of the action over ${\bf B}$, is:
 \begin{equation}
\mu_z = \gamma e^x_1 e^y_2  \sigma_z  \,,
 \label{MagneticMoment}
\end{equation}

Let us make the following transformation of the tetrad fields:
 \begin{equation}
e^x_1 \rightarrow  - e^x_1 \,\,, \,\,   e^y_2 \rightarrow  e^y_2    \,.
 \label{Transformation}
\end{equation}
Then the electron magnetic moment changes sign with respect to the spin:
 \begin{equation}
\mu_z  \rightarrow-  \gamma e^x_1 e^y_2  \sigma_z    \,,
 \label{Anomalous}
\end{equation}

The Zeeman energy $\mu_zB_z$ also changes sign after transformation  (\ref{Transformation}). The Zeeman energy of the electron is gravitating as any other energy. That is why the gravitational interaction of the body, in which the electrons are polarized  in the external magnetic field, with the other body, where electrons are not polarized, will be different for different sign of $e^x_1$. The Newton gravitational attraction will be increased after the change of sign. The original value of the gravitational interaction will be restored only after the electrons in the body will finally reorient their spins to reduce the Zeeman energy. 
    
This is the example of the symmetry breaking in the vacuum, which is caused by the existence of tetrads: vacua with different signs of the tetrad elements are not equivalent. 

There are the other examples related to spin, in particular related to the rotating black hole. 
If the black hole is formed by the spin-polarized matter, then the transformation (\ref{Transformation}) transforms the black hole to  the black hole with the opposite direction of rotation.

Also it is not excluded the existence of the the spin current in the vacuum in the presence of electric field,\cite{Chu2021}
which is analogous to spin supercurrent in $^3$He-B.\cite{Mineev1992}

\section{Weyl fermions and multiple tetrads}
\label{MultipleTetrads}

Let us now exploit the fact that the Weyl fermions are more fundamental than Dirac fermions. The Dirac are the secondary composite objects, which appear in the electroweak phase transition. Since the Weyl fermions are primary objects,  the action for gravity interacting with fermions should be written in terms of the Weyl fermions.  The interaction of gravity with  left and right fermions and with their antiparticles requires 4 different tetrads $e^\mu_{aA}$, $A=1,2,3,4 $: 
 \begin{equation}
{\cal L}_A= \Psi^+_A e^\mu_{aA} p_\mu \sigma^a  \Psi_A \,\,, \,\,  A=1,2,3,4  \,\,, \,\,  \sigma^a =(1, {\boldsymbol\sigma}) \,,
 \label{Weyl}
\end{equation} 
where ${\boldsymbol\sigma}$ are the Pauli matrices.
These four tetrads can be obtained from each other by $P_s$ and $T_s$ transformations. Choosing one of the 4 tetrads as the reference frame, for example the tetrad $e^\mu_{a{\rm R}}$ for right particles, one obtains the other three tetrads:
 \begin{equation}
e^\mu_{a{\rm R}}  \equiv e^\mu_{a}  \,\,, \,\, e^\mu_{a{\rm L}}  = P_s e^\mu_{a}\,\,, \,\,e^\mu_{a\bar{\rm R}} = P_sT_s e^\mu_{a} \,\,, \,\, e^\mu_{a\bar{\rm L}}  = T_s e^\mu_{a}
\,.
 \label{Weyl4}
\end{equation} 
This demonstrates the breaking of discrete symmetries $P_s$,  $T_s$  and $P_sT_s$ in the Minkowski vacuum.

Discussion of different tetrads for left and right Weyl fermions in Weyl materials can be found in Ref. \cite{NissinenVolovik2018b}.
The topological invariants for different types of the Weyl points are discussed in Ref.\cite{Zubkov2021}.

In our approach, four tetrads in Eq.(\ref{Weyl4}) are connected by symmetry operations, which means that the corresponding 
metric field is unique:
\begin{equation}
 g_{\mu\nu}^L \equiv g_{\mu\nu}^{\rm R}\equiv g_{\mu\nu} \,.
\label{CommonlMetric}
\end{equation}

\section{Possible bi-metric gravity from multiple tetrads}
\label{bimetric}

At low energies (well below the Planck scale) the original  symmetry between the fermions is violated. For example, in the Standard Model the symmetry between the left and right fermions is violated.  Could this lead to the violation of Eq.(\ref{CommonlMetric}), i.e. to the splitting of the metric field?
The bi-metricity does not take place, if on the fundamental level the metric is unique. However, one can try the scenarios, when the broken symmetry causes the small splitting between the metrics experienced by left and right fermions.  In this case this could lead to the bi-metric gravity at low energy.

Let us for example consider the scenario, in which the metric fields of different fermions are equal due to terms in action describing  diffeomorphism invariant interaction between the tetrads (or between the metrics, $U(g_{\mu\nu}^L ,g_{\mu\nu}^{\rm R})$), and the minimum of this potential $U$ corresponds to $g_{\mu\nu}^L = g_{\mu\nu}^{\rm R}$. 
When the symmetry between the left and right fermions is violated, the potential $U$ has corrections, $U=U_{\rm min} + \delta U$, where $\delta U$ contains the diffeomorphism invariant  corrections of first and second order in perturbation $g_{\mu\nu}^{\rm L} - g_{\mu\nu}^{\rm R}$:
\begin{equation}
\delta U \propto 
\epsilon  (g_{\mu\nu}^{\rm L} - g_{\mu\nu}^{\rm R})  (g^{\mu\nu {\rm L}} + g^{\mu\nu {\rm R}})
+ (g_{\mu\nu}^{\rm L} - g_{\mu\nu}^{\rm R})  (g^{\mu\nu {\rm L}} - g^{\mu\nu {\rm R}}) ,
\label{Action}
\end{equation} 
where $\epsilon \ll 1$. The metric fields for left and right fermions become slightly different:
\begin{equation}
g^{\mu\nu {\rm L}} - g^{\mu\nu {\rm R}} = -\epsilon g^{\mu\nu}\,.
\label{Action2}
\end{equation}
The $\epsilon$ correction to the potential comes from the integration over fermions. If the left and right fermions are different, the result will be different. This would mean that the Newton constant of gravitational interaction between fermions depends on their chiralities: there are 3 gravitational couplings, $G_ {\rm RR},G_{\rm LL},G_{\rm LR}$, i.e. the broken parity may lead to breaking of equivalence principle.

Different metrics for the left-handed and right-handed fermions has been also considered by Chadha and Nielsen,\cite{ChadhaNielsen1983} see also the so-called chiral gravity,\cite{Nibbelink2005}  and the bimetric theory for different particles in Ref.\cite{BernardBlanchet2015}. Although, the developing of such theories often suffers the ghost problem (Boulware-Deser ghost instability), see e.g. Ref.\cite{BernardBlanchetHeisenberg2015}, it is not excluded that due to the small value of the splitting in Eq.(\ref{Action2}), this problem can be avoided. Anyway, the possible scenario of emergent bi-metricity requires the detailed study.

\section{Tetrads and cosmological constant problem}
\label{CC}

Finally let us mention that the approach based on tetrads may have direct relation to the solution of the cosmological constant problem.
The original approach developed in Ref. \cite{KlinkhamerVolovik2008a,KlinkhamerVolovik2008}, which leads to the nullification of the cosmological constant in the Minkowski vacuum, was based on the nonlinear extension of the Hawking consideration of the vacuum variable as the four-form gauge field. Here we show that the role of the vacuum variable can be played by the tetrad determinant, $e\equiv {\rm det} e^a_\mu$. It is also the 4-form, but it is more natural for Standard Model than the 4-form gauge field. The simplest Einstein action in terms of  the tetrad determinant is:
\begin{equation}
 S=- \int_{\mathbb{R}^4}
\,d^4x\, e\,\left(\epsilon(e)+\frac{R}{16\pi G(e)} +\mathcal{L}^\text{M}(e,\psi)\right) \,.
\label{EinsteinAction}
\end{equation}
This action is invariant under coordinate transformations preserving the 4-volume. But it is not the unimodular gravity where $e={\rm const}$. Similar generalization of unimodular gravity has been considered in Ref. \cite{Klinkhamer2017}.

If one ignores the dependence of the matter action $\mathcal{L}$ and gravitational coupling $G$ on $e$, one obtains the Einstein equations with  the cosmological term $\rho_{\rm vac}(e) g_{\mu\nu}$, where the vacuum energy density is
\begin{equation}
 \rho_{\rm vac}(e) =\frac{\delta S}{\delta e}= \epsilon(e) + e\frac{d\epsilon(e)}{de} \,.
\label{VacEnergy}
\end{equation}
The fully equilibrium Minkowski vacuum (without matter) corresponds to the extremum of the action: 
\begin{equation}
\frac{\delta S}{\delta e}=0 \,.
\label{minimum}
\end{equation}
According to Eq.(\ref{VacEnergy}) the equilibrium value  $e=e_{\rm eq}$ in Minkowski vacuum is determined by
equation
\begin{equation}
\rho_{\rm vac}(e_{\rm eq})=0 \,,
\label{zero}
\end{equation}
which in turn corresponds to the nullification of that vacuum energy, which enters Einstein equations.
In this approach the cosmological constant in the equilibrium Minkowski vacuum is zero without fine tuning: the Planck scale quantity 
$\epsilon(e)$ is compensated by $e \,d\epsilon/de$.

Till now we considered positive determinant, $e>0$. But this can be extended to negative $e$. In this case one has two equilibrium states with opposite signs of $e$. The degenerate vacuua are connected by the $PT$ symmetry operation.\cite{Vergeles2021} That is why the appearance of nonzero $e$ corresponds to the symmetry breaking transition, at which the $PT$ symmetry is broken\cite{Vergeles2021} (together with other symmetries). The scale invariance is also violated in this transition. 

The simplest example is
\begin{equation}
e \epsilon(e)=m^4  e\left( \frac{1}{5}\frac{e^4}{e^4_{\rm eq}}  -1 \right) \,\,,\,\, \rho_{\rm vac}(e) = m^4 \left( \frac{e^4}{e^4_{\rm eq}}  -1 \right)  \,,
\label{example_e}
\end{equation}
where the scale $m$ could be the Planck energy scale.

\section{Extension to thermodynamics}
\label{Thermodynamics}

One can extend the determinant $e$ to the imaginary values. This corresponds to Euclidean  metric in the 4D space (in condensed matter the transition between Minkowski and Euclidean metric for Goldstone  modes see in Ref. \cite{NissinenVolovik2017}). On the imaginary axis  $e=i\beta$ the quantity $\beta$ becomes the inverse temperature in the 4+0 thermodynamics.  The path integral in quantum mechanics  is formally identical to the partition function $e^{-  {\cal S}}$ of a statistical mechanical system, with Hamiltonian $\epsilon(\beta)$ equal to the free energy: 
 \begin{equation}
 {\cal S}=  \int_{\rm Eu}
\,d^4x\, \beta\,\left(\epsilon(\beta)+\frac{R}{16\pi G(\beta)} +\mathcal{L}^\text{M}(\beta,\psi)\right) \,.
\label{EinsteinAction2}
\end{equation}

Then the vacuum energy density  becomes:
\begin{equation}
 \Lambda\equiv \rho_{\rm vac} (T)=  \epsilon - T\frac{d\epsilon}{dT}=  \epsilon+ \beta\frac{d\epsilon(\beta)}{d\beta} =  \frac{d\left(\beta \epsilon(\beta)\right)}{d\beta}= \frac{d{\cal S}(\beta)}{d\beta} .
\label{Lambda-beta}
\end{equation}
The extremum of partition function suggests that in thermodynamic equilibrium and in the absence of external pressure the cosmological constant is nullified:
\begin{equation}
  \Lambda_{\rm equilibrium}=\frac{d{\cal S}_{\rm vac}(\beta)}{d\beta}=0
\,.
\label{LambdaZero}
\end{equation}
Eq.(\ref{LambdaZero}) determines the equilibrium value $T_{\rm eq}$ of temperature (and correspondingly the non-zero equilibrium value $e=e_{\rm eq}$ in Minkowski vacuum). 
Example in Eq.(\ref{example_e}) extended to imaginary $e=i\beta$ gives
\begin{equation}
\beta \epsilon(\beta)=m^4  \beta\left( \frac{1}{5}\frac{\beta^4}{\beta^4_{\rm eq}}  -1 \right) 
\,\,,\,\, \rho_{\rm vac}(\beta) = m^4\left( \frac{\beta^4}{\beta^4_{\rm eq}}  -1 \right) .
\label{example_beta}
\end{equation}

So, the non-zero temperature of the 4+0 system gives rise to gravity in the 3+1 system. This is another connection between gravity and thermodynamics in addition to that suggested by Jacobson.\cite{Jacobson1995} The original 4D quantum vacuum has $\beta=0$, and this infinite temperature corresponds to complete chaos, where gravity is absent. The finite $T_{\rm eq}$ gives rise to the Einstein action, and is responsible for the equilibrium Minkowski vacuum with equilibrium value of the Newton constant $G(e_{\rm eq})$ and zero value of the cosmological constant. It is also related to the existence of the elementary (Planck scale) 4-volume. 

While in thermodynamics the nullification of the cosmological constant in the full equilibrium is natural,
in dynamics the de Sitter attractors may prevent the approach to the equilibrium state of the quantum vacuum.\cite{KlinkhamerVolovik2008}

\section{Conclusion}

We considered two scenarios of the formation of the combined symmetry describing the tetrad gravity interacting with fermions: the scenario of emergent symmetry suggested by analogy with the chiral superfluid $^3$He-A and the scenario of symmetry breaking, which is suggested by the symmetry breaking pattern in topological superfluid $^3$He-B. In the first one the topology of the Weyl points leads to the tetrads emerging in the vicinity of the Weyl point. In the second one the gravitational tetrads $e_a^\mu$ appear as the order parameter of the symmetry breaking phase transition, which is the bilinear  form of the fermionic fields.\cite{Diakonov2011,Diakonov2012,ObukhovHehl2012,Wetterich2003,Wetterich2012,Wetterich2013}  

Such broken symmetry phase transition gives rise to the degeneracy of the Minkowski  vacua, the particular example -- the broken $PT$ symmetry in the Minkowski vacuum -- was discussed  by Vergeles.\cite{Vergeles2021}   In this scenario the original trans-Planckian vacuum was invariant under coordinate transformations: the Lorentz $L_{\rm c}$ coordinate transformation and the space and time reversal transformations, $P_{\rm c} {\bf r}=-{\bf r}$ and $T_{\rm c}t =-t$.  The tetrad order parameter, which leads to formation of gravity, spontaneously breaks these symmetries.
The degenerate quantum vacua  obtained  after the symmetry breaking loose the invariance under the pure coordinate transformations. The broken symmetry vacua obey only the combined symmetries: $P=P_{\rm c}P_{\rm s}$ and $T=T_{\rm c}T_{\rm s}$ symmetries as well as combined Lorentz symmetry $L=L_{\rm c}L_{\rm s}$, where the coordinate transformation  is accompanied by the corresponding rotation or reflection of the order parameter -- the tetrad -- in spin space.  

Such mechanism of the formation of gravity leads to many interesting effects, including the topological defects formed by this symmetry breaking due to Kibble-Zurek mechanism, and to the  four tetrad fields for four types of Weyl  fermions: for left-handed fermions, for right-handed fermions and for their corresponding antiparticles. In principle, in some scenarios the parity violation in Standard Model may lead to the breaking of equivalence principle.

The discussed symmetry breaking, $L_{\rm c}  \times L_{\rm s}  \rightarrow L= L_{\rm c}   L_{\rm s}$,  $P_{\rm c}  \times P_{\rm s}  \rightarrow P= P_{\rm c}   P_{\rm s}$ and  $T_{\rm c}  \times T_{\rm s}  \rightarrow T= T_{\rm c}   T_{\rm s}$, probably takes place at the Planck or higher energy scale. At the lower energy scale, where the Standard Model operates, the combined symmetry of the degenerate vacuum is further reduced by the breaking of the discrete $P$ and $T$ symmetries.  Which of these symmetry breaking  transitions is responsible for the observed baryon asymmetry of the Universe and for the strong CP problem, is an open question raised in Ref.\cite{Vergeles2021}.

Finally we considered the application of the tetrad fields for the solution of the cosmological constant problem.

{\bf Acknowledgements}. 
I thank S. Vergeles and M. Zubkov for discussions. This work has been supported by the European Research Council (ERC) under the European Union's Horizon 2020 research and innovation programme (Grant Agreement No. 694248).


\begin{thebibliography}{99}


\bibitem{Volovik2021}
G.E. Volovik,
$^3$He Universe 2020,
J. Low Temp. Phys. {\bf 202}, 11--28 (2021),
https://doi.org/10.1007/s10909-020-02538-8,
arXiv:2008.04682.

\bibitem{Volovik2003}
G.E. Volovik,
The Universe in a Helium Droplet,
Clarendon Press,  Oxford (2003)

\bibitem{VollhardtWolfle}
D. Vollhardt and P. Wölfle, 
The Superfluid Phases of Helium 3 (Taylor \& Francis, London, 1990).

\bibitem{Vergeles2021}
S.N. Vergeles,
A note on the vacuum structure of lattice Euclidean quantum gravity: birth of macroscopic space-time and $PT$-symmetry breaking,
 Class. Quantum Grav. {\bf 38}, 085022 (2021).

\bibitem{Volovik1990}
G.E. Volovik,
Superfluid $^3$He-B and gravity,
Physica B {\bf 162}, 222--230 (1990).

\bibitem{Leggett1973}
A.J. Leggett, 
NMR lineshifts and spontaneously broken spin-orbit symmetry. I. General concepts,
 J. Phys. C {\bf 6}, 3187 (1973).

\bibitem{NissinenVolovik2018}
J. Nissinen and G.E. Volovik,
Dimensional crossover of effective orbital dynamics in polar distorted  $^3$He-A: Transitions to anti-spacetime,
Phys. Rev. D {\bf 97}, 025018  (2018),
arXiv:1710.07616.

\bibitem{Autti2020}
S. Autti, J. T. M\"akinen, J. Rysti, G. E. Volovik, V. V. Zavjalov, V. B. Eltsov,
 Exceeding the Landau speed limit with topological Bogoliubov Fermi surfaces,
Physical Review Research {\bf 2}, 033013 (2020),
 arXiv:2002.11492.
 
 \bibitem{SilaevVolovik2010}
M.A. Silaev and G.E. Volovik,
Topological superfluid $^3$He-B: fermion zero modes on interfaces and in the vortex core,
J. Low Temp. Phys, {\bf 161},  460--473 (2010),
arXiv:1005.4672. 

\bibitem{Diakonov2011}
D. Diakonov,
Towards lattice-regularized Quantum Gravity,
arXiv:1109.0091 (2011).

\bibitem{Diakonov2012}
A.A. Vladimirov, D. Diakonov,
Phase transitions in spinor quantum gravity on a lattice,
Phys. Rev. D {\bf 86}, 104019 (2012). 

 \bibitem{ObukhovHehl2012}
Y.N. Obukhov and F.W. Hehl,
Extended Einstein–Cartan theory a la Diakonov: The field equations,
Phys. Lett. B {\bf 713}, 321--325 (2012).

\bibitem{Wetterich2003}
A. Hebecker and C. Wetterich,
Spinor gravity,
Physics Letters B {\bf 574} 269--275, (2003).

\bibitem{Wetterich2012}
C. Wetterich,
Universality of geometry,
Physics Letters B {\bf 712}, 126--131 (2012).

\bibitem{Wetterich2013}
D. Sexty and C. Wetterich,
Emergent gravity in two dimensions,
Nuclear Physics B {\bf 867}, 290--329 (2013).

\bibitem{SalomaaVolovik1988}
M.M. Salomaa,  G.E. Volovik, 
Cosmiclike domain walls in superfluid $^3$He-B: Instantons and diabolical points in (${\bf k}$, ${\bf r}$) space, 
Phys. Rev. B {\bf 37}, 9298--9311 (1988).

\bibitem{Kibble1982}
T. W. B. Kibble, G. Lazarides, and Q. Shafi,
Walls bounded by strings,
Phys. Rev. D {\bf 26}, 435--439 (1982);
Strings in $SO(10)$,
Phys. Lett. B {\bf 113}, 237--239 (1982).

\bibitem{Makinen2019}
 J.T. M\"akinen, V.V. Dmitriev, J. Nissinen, J. Rysti, G.E. Volovik, A.N. Yudin, K. Zhang, V.B. Eltsov,
Half-quantum vortices and walls bounded by strings in the polar-distorted phases of topological superfluid $^3$He,
Nat. Comm. {\bf 10}, 237 (2019), 
arXiv:1807.04328.

\bibitem{Kuang2020}
G. E. Volovik, K. Zhang,
String monopoles, string walls, vortex-skyrmions and nexus objects in polar distorted B-phase of $^3$He,
Physical Review Research {\bf 2}, 023263 (2020),
arXiv:2002.07578.

\bibitem{SalomaaVolovik1987}
M.M. Salomaa,  G.E. Volovik, 
Quantized Vortices in superfluid $^3$He,
Rev. Mod. Phys. {\bf 59}, 533--613 (1987).

\bibitem{Kondo1992}
Y. Kondo,  J.S. Korhonen, M. Krusius, V.V. Dmitriev, E.V. Thuneberg and  G.E. Volovik, 
Combined spin-mass vortex with soliton tail in superfluid  $^3$He-B,
Phys. Rev. Lett. {\bf 68}, 3331 (1992).

\bibitem{Jacobson} 
T. Jacobson, private communications.

\bibitem{HansonRegge} 
A.J. Hanson and T. Regge, 
Torsion and quantum gravity, 
in: Proceedings of the Integrative Conference on Group Theory
and Mathematical Physics, University of Texas at Austin, 1978.

\bibitem{AuriaRegge} 
R. d'Auria and T. Regge, 
Gravity theories with asymptotically flat instantons,
Nucl. Phys. {\bf B~195}, 308 (1982).

\bibitem{Sakharov1980} 
A.D. Sakharov,
Cosmological models of the Universe with reversal of time's arrow,
JETP {\bf 52}, 349--351 (1980).

\bibitem{Turok2018}
L. Boyle, K. Finn and N. Turok, 
CPT-symmetric universe,
Phys. Rev. Lett. {\bf 121}, 251301 (2018).

\bibitem{Turok2021}
L. Boyle and N. Turok, 
Two-sheeted Universe, analyticity and the arrow of time,
arXiv:2109.06204.

\bibitem{Chu2021}
Chong-Sun Chu and Chun-Hei Leung,
Induced quantized spin current in vacuum,
Phys. Rev. Lett. {\bf 127}, 111601 (2021).

\bibitem{Mineev1992}
 V.P. Mineev and G.E. Volovik, 
 Electric dipole moment and spin supercurrent in superfluid $^3$He-B,   
J. Low Temp. Phys. {\bf 89 }, 823--830 (1992).

\bibitem{NissinenVolovik2018b}
J. Nissinen and G.E. Volovik,
Tetrads in solids: from elasticity theory to topological quantum Hall systems and Weyl fermions,
ZhETF {\bf 154},   1051--1056 (2018),
JETP {\bf 127}, 948--957 (2018),
arXiv:1803.09234.

\bibitem{Zubkov2021}
M.A. Zubkov,
Classification of emergent Weyl spinors in multi-fermion systems,
Pis'ma ZhETF  {\bf 113}, 448--449 (2021);
JETP Lett. {\bf 113}, 445--453 (2021),
arXiv:2102.00964.

\bibitem{ChadhaNielsen1983}
S. Chadha and H.B. Nielsen,
Lorrentz invariance as a low-energy phenomenon,
Nucl. Phys. B {\bf 217}, 125--144 (1983).

\bibitem{Nibbelink2005}
Stefan Groot Nibbelink and Marco Peloso,
Chiral gravity as a covariant formulation of massive gravity,
Class. Quantum Grav. {\bf 22} 1313--1327 (2005).

\bibitem{BernardBlanchet2015}
L. Bernard and L. Blanchet,
Phenomenology of dark matter via a bimetric extension of general relativity,
Phys. Rev. D {\bf 91}, 103536 (2015).

\bibitem{BernardBlanchetHeisenberg2015}
L. Bernard, L. Blanchet and L. Heisenberg, 
Bimetric gravity and dark matter,
in: Proceedings, 50th Rencontres de Moriond Gravitation: 100 years after GR : La Thuile, Italy, March 21-28, 2015, 43--52,
arXiv:1507.02802

\bibitem{KlinkhamerVolovik2008a}
F.R. Klinkhamer and G.E. Volovik,  
Self-tuning vacuum variable and cosmological constant, 
Phys. Rev. D {\bf 77}, 085015 (2008).
 
\bibitem{KlinkhamerVolovik2008}
F.R. Klinkhamer and G.E. Volovik,  
Dynamic vacuum variable and equilibrium approach in cosmology,   
Phys. Rev. D {\bf 78}, 063528 (2008).

\bibitem{Klinkhamer2017}
F. R. Klinkhamer,
A generalization of unimodular gravity with vacuum-matter energy exchange,
International Journal of Modern Physics D {\bf 26}, 1750006 (2017).

\bibitem{NissinenVolovik2017} 
J. Nissinen and G.E. Volovik,
Effective Minkowski-to-Euclidean signature change of the magnon BEC pseudo-Goldstone mode in polar $^3$He,
Pis'ma ZhETF {\bf 106}, 220--221 (2017), 
JETP Lett. {\bf 106},  234--241 (2017),
arXiv:1707.00905.

\bibitem{Jacobson1995}
T. Jacobson,
Thermodynamics of Spacetime: The Einstein Equation,
Phys. Rev. Lett. {\bf 75}, 1260 (1995).

\end{thebibliography}
\end{document}